\documentclass{ws-procs975x65}

\begin{document}

\title{THERMAL ASPECTS IN CURVED METRICS}

\author{GIOVANNI ACQUAVIVA$^*$}

\address{Department of Physics, University of Trento,\\
Trento, Italy\\
$^*$E-mail: acquaviva@science.unitn.it}

\begin{abstract}
In this paper we describe two approaches that allow to calculate some thermal features as perceived by different observers in curved spacetimes: the tunnelling method and the Unruh-DeWitt detector.  The tunnelling phenomenon is a semi-classical approach to the issue of Hawking radiation and allows a straightforward calculation of the horizon temperature in a plethora of scenarios; the Unruh-DeWitt model relies instead on a quantum field-theoretical approach and (whenever possible) gives a more exact answer in terms of transition rates between energy levels of an idealized detector.
\end{abstract}


\bodymatter

\section{The tunnelling method(s)}

Both the \textit{null-geodesic} method (by Kraus, Parikh and Wilczek in Refs.~\refcite{kra,par}) and the \textit{Hamilton-Jacobi} method (by Padmanabhan and collaborators in Ref.~\refcite{sri}) rely on the calculation of the classical action $S$ of a particle along a trajectory crossing the horizon from the trapped region towards the observer's region.  In a WKB approximation, the tunnelling probability rate is given by
\begin{equation}\label{im}
 \Gamma_{em} \simeq e^{-2\, \text{Im}(S)}
\end{equation}
so it is clear that a non-vanishing probability of emission corresponds to the presence of an imaginary contribution coming from the action along the trajectory considered.  Moreover, comparing the expression Eq. (\ref{im}) with the Boltzmann factor, one should be able to relate the imaginary part of the action to the quantity $\beta\, \omega$, identifying in this way the temperature of the emitted radiation.  Restricting the attention to the Hamilton-Jacobi method, the calculation of emission rate can be summarized in the following steps:

\begin{arabiclist}[3]
 \item assume that the tunnelling particle's action $S$ satisfies the relativistic Hamilton-Jacobi equation
 \begin{equation}
  g^{\mu \nu} \partial_{\mu}S\, \partial_{\nu}S\, + m^2 = 0
 \end{equation}

 \item reconstruct the whole action, starting from the symmetries of the problem; the integration is carried along an oriented, null, curve $\gamma$ with at least one point on the horizon
 \begin{equation}
  S = \int_{\gamma}\, dx^{\mu}\, \partial_{\mu}S
 \end{equation}

 \item perform a near-horizon approximation and regularize the divergence in the integral according to Feynman's prescription: the solution of the integral has in general a non-vanishing imaginary part.
\end{arabiclist}
The present author and his collaborators in Ref.~\refcite{acqua} reviewed the tunnelling approach in a wide variety of situations.  The Kodama-Hayward theoretical results (see Refs.~\refcite{kod,hay}) have been one of the main ingredients that allowed to extend the method to dynamical scenarios and enabled to express the combination $\beta\, \omega$ as an invariant quantity:
\begin{equation}
 \Gamma_{em} = \Gamma_{abs}\, \exp\left( - \frac{2 \pi \omega_H}{\kappa_H} \right)
\end{equation}
where $\omega_H$ is the tunnelling particle's energy (conserved with respect to the Kodama vector) and $\kappa_H$ is Hayward's surface gravity.  In this way one can identify an invariant temperature $T_H = \frac{\kappa_H}{2\pi}$.\\
The thorough analysis in Ref.~\refcite{acqua} resulted in the following achievements:
\begin{itemlist}
 \item a solid basis for the covariance of the method has been given; the question whether horizons have a temperature finds here an invariant answer.
 \item formal equivalence of the two approaches (null-geodesic and Hamilton-Jacobi) holds at least in stationary cases;
 \item the method provides an invariant and consistent answer in a variety of situations (higher-dimensional solutions, Taub and Taub-NUT solutions, decay of unstable particles, emission from cosmological horizons and naked singularities).
\end{itemlist}

\section{Unurh-DeWitt detectors}

The Unruh-DeWitt detector (see Refs.~\refcite{unruh,dewitt}) provides a more exact answer to questions regarding the particle content of a field in a curved metric and its thermal features for different observers.  In Ref.~\refcite{udw} the authors consider a conformally flat 4-dimensional metric, a massless scalar field conformally coupled to the metric and a two-level quantum system coupled the the scalar field.  The idea is to calculate the probability for the absorption of a scalar quantum and the consequent excitation of the two-level system through the \emph{transition rate}
\begin{equation}\label{resp}
 \frac{dF}{d\tau} = \frac{1}{2\pi^2} \int_0^{\infty}\ \cos\left(E\, s\right)\, \left( \frac{1}{\sigma^2(\tau,s)} + \frac{1}{s^2} \right)\, ds\ - \frac{1}{2\pi^2} \int_{\Delta\tau}^{\infty} \frac{\cos \left( E\, s \right)}{\sigma^2(\tau,s)}
\end{equation}
where $E$ is the energy gap of the detector and $s$ is the duration of the detection (see Ref.~\refcite{udw} for details on the construction of Eq. (\ref{resp})).  The second integral is the \emph{finite-time contribution}.  The bulk of the information about the transition rate comes from the geodesic distance between the ``switching on'' and ``switching off'' events, evaluated along a fixed trajectory $x(\tau)$
\begin{equation}
 \sigma^2(\tau,s) = a(\tau) a(\tau-s)\, \left[ x(\tau) - x(\tau-s) \right]^2
\end{equation}
where the $a(t)$ is the conformal factor.  The inverse of $\sigma^2$ is proportional to the positive frequency Wightman function.\\
In the paper the authors analyzed the \emph{Schwarzschild black hole} and the \emph{de Sitter model}.  The detector has been placed on a Kodama trajectory, which means that it sits at fixed areal radius.  Both cases can be treated in the same way, because the function $\sigma^2$ can be written in general
\begin{equation}\label{sigma}
 \sigma^2(s) = - \frac{4 V}{\kappa^2} \sinh^2\left( \frac{\kappa}{2 \sqrt{V}}\, s \right)
\end{equation}
where $\kappa$ is the surface gravity and $\sqrt{V} = \sqrt{-g_{00}}$.  A Wightman function which, as in Eq. (\ref{sigma}), is stationary and periodic in imaginary time is called ``thermal'' because when Fourier-transformed gives a Planckian transition spectrum.  In our case, calculating both the stationary and the finite-time contributions, the transition rate reads
\begin{eqnarray}
 \frac{dF}{d\tau}&=&\frac{1}{2\pi} \frac{E}{\exp\left( \frac{2\pi \sqrt{V} E}{\kappa} \right) - 1}\ +\nonumber \\
  && +\, \frac{E}{2 \pi^2} \sum_{n=1}^{\infty}\, \frac{n\, e^{-n \kappa \Delta\tau / \sqrt{V}}}{n^2 + V E^2/\kappa^2} \left( \frac{\kappa}{\sqrt{V} E} \cos(E \Delta\tau) - \sin(E \Delta\tau) \right)
\end{eqnarray}

\section{Conclusions}

As regards the tunnelling method, it has been shown that the formalism gives an invariant answer and allows extensions to more general black hole horizons in various dimensions as well as cosmological horizons and naked singularities.  Moreover, the extension to dynamical spacetimes has been carried out: in this framework the radiation seems to originate near the local trapping horizon, not the global event horizon (which has well-known teleological issues).\\
The Unruh-DeWitt detector constitutes a more exact approach to the Unruh-Hawking effect, relying on a quantum field-theoretical calculation. In stationary cases the response function of the detector is shown to be thermal with temperature given by the surface gravity, just as in the tunnelling approach.  The generalization to non-stationary situations gives rise to problems in the analytical resolution and in general, when the background is time-dependent, the thermal interpretation seems lost.

\end{document}